\documentclass[11pt]{article}

\usepackage[utf8]{inputenc}
\usepackage[colorlinks=true, allcolors=blue]{hyperref}
\usepackage[margin=1 in]{geometry}
\usepackage{graphicx}
\usepackage{amsmath, amssymb, mathtools, mathrsfs, dsfont, amsthm}
\usepackage{enumerate}
\usepackage{fancyhdr}
\usepackage{authblk}
\usepackage{url}
\usepackage{lipsum}
\usepackage{tensor}
\usepackage{comment}
\graphicspath{/fig}
\usepackage{subcaption}

\DeclareMathOperator{\Tr}{Tr}
\DeclareMathOperator{\sech}{sech}
\DeclareMathOperator{\csch}{csch}

\theoremstyle{definition}

\theoremstyle{remark}

\theoremstyle{remark}

\title{\textbf{Islands and mixed states in closed universes}}
\author{Seamus Fallows\footnote{seamus.fallows@durham.ac.uk} }
\author{Simon F. Ross\footnote{s.f.ross@durham.ac.uk}}
\affil{\textit{Centre for Particle Theory, Department of Mathematical Sciences Durham University, South Road, Durham DH1 3LE, U.K.}}
\date{\today}

\begin{document}
\maketitle
\begin{abstract}
We investigate the appearance of islands when a closed universe with gravity is entangled with a non-gravitating quantum system. We use braneworlds in three-dimensional multiboundary wormhole geometries as a model to explore what happens when the non-gravitating system has several components. The braneworld can be either completely contained in the entanglement wedge of one of the non-gravitating systems or split between them. In the former case, entanglement with the other system leads to a mixed state in the closed universe, unlike in simpler setups with a single quantum system, where the closed universe was necessarily in a pure state. We show that the entropy of this mixed state is bounded by half of the coarse-grained entropy of the effective theory on the braneworld. 
\end{abstract}
\newpage
\tableofcontents 

\section{Introduction}

The Ryu-Takayanagi proposal \cite{Ryu:2006bv} and its generalizations \cite{Hubeny:2007xt,Faulkner:2013ana,Engelhardt:2014gca} have given us important insights into the description of spacetime in holographic theories of quantum gravity. This has recently been extended by the discovery of the island phenomenon \cite{Penington:2019npb,Almheiri:2019psf,Almheiri:2019hni}: when we consider entangling the holographic theory with another quantum system, the part of the spacetime described by the holographic dual can be bounded by a quantum extremal surface \cite{Engelhardt:2014gca}, and the spacetime region beyond this surface (the island) is encoded in the other system. These ideas were initially developed for holographic theories, but they can be derived from the Euclidean gravitational path integral \cite{Penington:2019kki,Almheiri:2019qdq}, so they are believed to apply more generally. 

Consider a conventional quantum mechanical system (which could be a quantum field theory, or a simpler system, such as a spin chain) which is entangled with a gravitational system. The island rule is that given a semiclassical, effective description of the gravitational system, the {\it fine-grained}  entropy of some subsystem $A$ in the quantum system is given by 
\begin{equation} \label{si}
S(A) = \mathrm{min}_{\mathcal I}\  \mathrm{ext}_{\mathcal I}\  S_{\mathrm{gen}} (A \cup \mathcal I),   
\end{equation}
where $\mathcal I$ is the {\it island}, some spatial subregion of the spacetime the gravitational system lives in, and $S_{\mathrm{gen}}$ is the generalised entropy. If the semiclassical theory is Einstein gravity coupled to matter, $S_{\mathrm{gen}} (A \cup \mathcal I) = \frac{A_{\partial \mathcal I}}{4G} + S_{\mathrm{eff}}(A  \cup \mathcal I)$, where $A_{\partial \mathcal I}$ is the area of the boundary of the island and $S_{\mathrm{eff}}$ is the von Neumann entropy of the effective semi-classical state of $A$ together with the fields in the island. The spatial subregion that extremises $S_{\mathrm{gen}}$ is called a quantum extremal island; the surface $\partial \mathcal I$ is the quantum extremal surface. If there are multiple islands in the spacetime, we choose the one which minimizes the entropy. 

We include an island in the spacetime if there is sufficient entanglement between $A$ and the spacetime in the semi-classical state to compensate for the large contribution to $S_{\mathrm{gen}}$ from the area term. Then $S(A) < S_{\mathrm{eff}}(A)$, and the true fine-grained entropy of $A$ calculated according to this prescription is smaller than the effective entropy. The derivation of this formula from the path integral  \cite{Penington:2019kki,Almheiri:2019qdq} tells us that the effective semi-classical state in the island region is encoded in $A$; semi-classically it looks like we have a seperate Hilbert space $\mathcal H_{\mathcal I}$ of the quantum fields on the island, but in fact this is encoded as a code subspace in the Hilbert space $\mathcal H_A$ of the quantum system $A$.

The path integral derivation suggests this prescription applies quite generally, and in particular we can consider its application to a closed universe. We do not have a good understanding of quantum gravity in closed universes, but we can still apply the island rule, as it only requires an understanding of the semi-classical state. In a closed universe $U$, if the gravitating system is entangled with $A$ but otherwise in a pure state, then we can take the island to be the whole universe \cite{Almheiri:2019hni,Chen:2020tes}.\footnote{For another perspective see \cite{Balasubramanian:2020coy,Balasubramanian:2020xqf}.} Then there is no boundary term, and $S_{\mathrm{gen}} (A \cup U) =  S_{\mathrm{eff}}(A  \cup U)$. As soon as there is any entanglement between the closed universe and $A$, the whole closed universe is encoded in $\mathcal H_A$. In section \ref{braneworld}, we review a simple doubly holographic model which illustrates this in a well-understood setting. Thus, entanglement allows us to recover some aspects of a theory of quantum gravity on a closed universe from this quantum system, in a way which generalizes the holographic story. 

This might seem surprising, as the semiclassical theory on $U$ could have a Hilbert space $\mathcal H_U$ which is much larger than $\mathcal H_A$.  However, we have so far considered only semi-classical states on $U$ which have some entanglement with $A$ but are otherwise pure. The purpose of the present paper is to explore the extension to cases where the closed universe is in a mixed state.  In the holographic context, even if individual pure states are encoded in a dual system, this encoding can break down when we consider mixtures of them; this is the essential issue underlying the original appearance of the islands in \cite{Penington:2019npb,Almheiri:2019psf}. To  fully understand encoding of the semiclassical Hilbert space $\mathcal H_U$ in  $\mathcal H_A$, we should ask how the encoding works when we consider mixed states in $\mathcal H_U$ with some entanglement with $A$. We can express this by considering a situation where $U$ is entangled both with $A$ and with some other quantum system $B$ which acts as a purifier for the mixed state in $\mathcal H_U$. There is then clearly a competition between entanglement with $A$ and entanglement with $B$, and we can have either $S_{\mathrm{eff}}(A  \cup U)< S_{\mathrm{eff}}(A)$ or $S_{\mathrm{eff}}(A  \cup U)> S_{\mathrm{eff}}(A)$; the whole universe is not always included in an island for $A$. 

The qualitative picture is very similar to the holographic story mentioned above: if we start with a situation where the spacetime is encoded in a single quantum system, and add some additional entanglement of the semi-classical state on the spacetime with another system, initially this describes a mixed state of the quantum fields on the spacetime. When the entanglement with the new system gets large enough, a new island appears, and the spacetime is encoded partially or wholly in this second system. We will explore the limitations this imposes on the entropy of the mixed state we can consider in the context of our simple model. 

We explore this issue in detail in a simple doubly holographic braneworld model, proposed in \cite{Cooper:2018cmb}. The original model, reviewed in section \ref{braneworld}, has a bulk spacetime with two boundaries, where one boundary is a dynamical brane and the other boundary is an asymptotically AdS boundary. We take both boundaries to be closed spaces, and they are connected by a spatial wormhole (Einstein-Rosen bridge) in the bulk. By integrating out the bulk spacetime, we can obtain a semi-classical description of this spacetime where we have a CFT coupled to gravity on the dynamical brane, and a second non-gravitating CFT on the asymptotic boundary, in an entangled state. Then according to the island formula, when we calculate the fine-grained entropy of the non-gravitating CFT, we include the whole of the other closed universe in an island. This corresponds simply to the bulk statement that the entanglement wedge for the CFT on the single asymptotically AdS boundary includes the whole of the bulk spacetime. Semi-classically we had an entangled state relating degrees of freedom on the brane and degrees of freedom on the boundary, but the microscopic theory is a field theory on the boundary; the semi-classical state is encoded in some subspace in this. 

One approach to studying a brane entangled with multiple systems would be to simply subdivide the non-gravitating boundary in this model into several regions. The appearance of islands associated with subregions in this model was indeed already explored in \cite{Cooper:2018cmb}. We instead consider an extension of this model with a brane and multiple asymptotic regions, dual to several copies of a CFT. These different boundaries are connected by a multiboundary wormhole. We will focus on a wormhole with three asymptotic regions, one of which we cut off with a dynamical brane. This has a couple of advantages: since the two boundaries aren't coupled, the entanglement between them is time-independent and free of ultraviolet divergences. The lengths of the horizons associated with the different asymptotic regions are also all independent parameters. 

We review the bulk wormhole geometry in section \ref{wormhole}. We introduce the brane in section \ref{islands}. In the semi-classical description, we have a CFT coupled to gravity on the brane, and non-gravitating CFTs on the asymptotic boundaries, in some entangled state $|\tilde \Sigma \rangle$. Microscopically, the whole spacetime, including the brane, is encoded in a state $|\Psi \rangle$ on the two asymptotic boundaries. To determine which of the CFTs the brane is encoded in, we determine the fine-grained entropies by comparing the different possible Ryu-Takayanagi (RT) surfaces in the bulk. The brane can be fully encoded in one of the two systems or partially in each, with a quantum extremal surface dividing the two regions. The latter occurs if the entanglement between the brane and the other systems is large enough to compensate for the boundary term in $S_{\mathrm{ gen}}$. 

We can treat one of the asymptotic boundaries as a reference system; tracing over this, we obtain a mixed state both semi-classically and microscopically. The semi-classical state will give a mixed state on the brane if it is entirely encoded in the other system, but still has some entanglement with the reference system. In this case we have a microscopic mixed state, part of whose entropy is associated with the semi-classical mixed state on the brane. When we consider multiboundary wormholes with long horizons, as in \cite{Marolf:2015vma}, we can split the microscopic entropy into a part associated semi-classically with entanglement between the reference system and the brane and a part associated with the entanglement between the reference system and the other system. We show that the entropy of the semi-classical mixed state on the brane has two bounds: it is always less than half the coarse-grained entropy of the fields on the brane in this semi-classical state (otherwise it would be favourable to have the brane entirely encoded in the reference system rather than the other boundary) and less than the contribution to the generalized entropy from boundaries of an island (otherwise it would be favourable to have an island on the brane). The simplicity of the latter bound is due to our model describing a particular kind of mixed state, where the reference system is entangled with a particular local region on the brane. We could certainly imagine entangling the brane with the reference system in more complicated ways, which could relax this bound. The first bound seems more universal. 

We conclude with a brief discussion and consideration of future directions in section \ref{disc}. 

\section{Braneworld model}
\label{braneworld} 

The model we consider was proposed in \cite{Cooper:2018cmb}; this was further developed in \cite{Antonini:2019qkt,VanRaamsdonk:2020tlr,VanRaamsdonk:2021qgv,Sully:2020pza}. It can be seen as an example of the doubly holographic setup of \cite{Almheiri:2019hni}, which was applied to island calculations in \cite{Chen:2020tes}. The idea is to consider an end of the world brane in AdS which has a closed universe cosmology as its worldvolume, and seek insight into the quantum theory of the closed universe from the holographic dual CFT.  

We will work with the simplest model of an end of the world brane, developed in  \cite{Takayanagi:2011zk,Fujita:2011fp}. We consider a three-dimensional locally AdS$_3$ bulk, dual to a two-dimensional CFT, with a constant-tension end of the world brane in the bulk, holographically dual to some one-dimensional boundary degrees of freedom coupled to the CFT$_2$. The bulk three-dimensional theory has action 
\begin{equation} \label{I3d} 
I = \frac{1}{16 \pi G} \int_M d^3 x \sqrt{-g} (R-2\Lambda) + \frac{1}{8\pi G} \int_{\partial M}  d^2 y \sqrt{-h}\, K - \frac{1}{8 \pi G} \int_{Q} d^2 y \sqrt{-h}\,  \frac{T}{\ell} 
\end{equation}
where $G$ is the three-dimensional Newton constant, $\Lambda = -\frac{1}{\ell^2}$ is a cosmological constant, $K$ is the trace of the extrinsic curvature and $T$ is the tension of the end of the world brane with worldvolume $Q$, which we take to be one component of the boundary $\partial M$ of the spacetime, the other component(s) corresponding to asymptotically AdS boundary(ies). We will work in units where the AdS scale $\ell=1$. The brane has a stress-energy tensor $8\pi G T_{ab}= -T  h_{ab}$, and the action implies that the boundary condition for the bulk metric at $Q$ is $K_{ab} - K h_{ab} = 8\pi G T_{ab} = - T h_{ab}$. 

\begin{figure}[ht]
\centering
 \includegraphics[width=.5\linewidth]{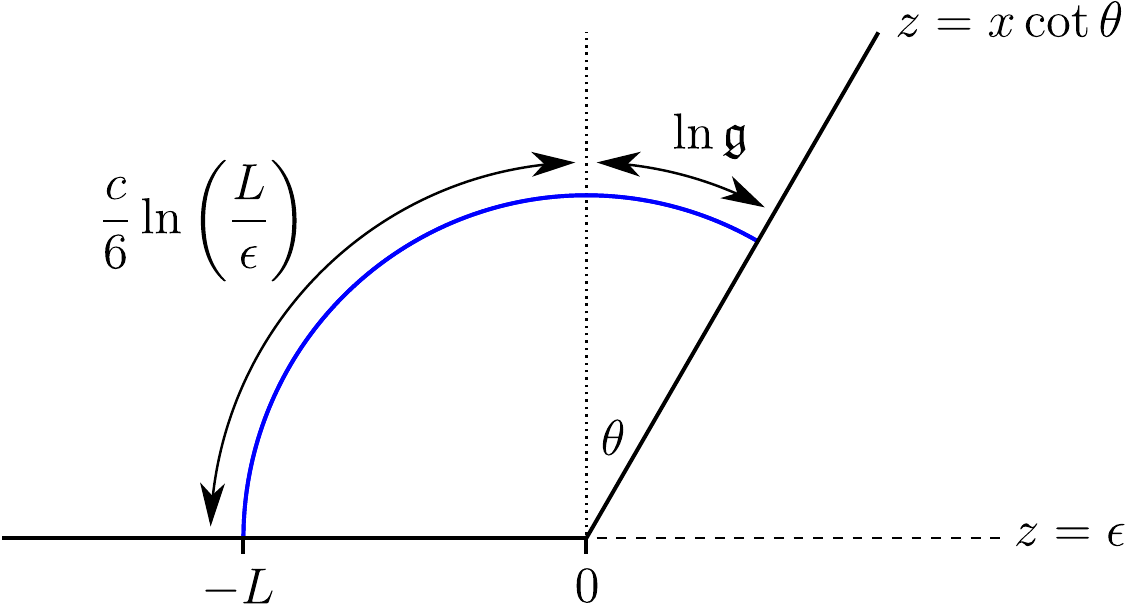}  
\caption{A constant-time slice of the geometry with an end of the world brane in Poincar\'e-AdS, showing the RT surface for a region $x \in (-L,0)$ on the boundary}
\label{fig:poinc}
\end{figure}

A simple solution of this theory is pure AdS$_3$ in Poincar\'e coordinates, 
\begin{equation}
ds^2 = \frac{-dt^2+dx^2 + dz^2}{z^2}, 
\end{equation}
with the end of the world brane $Q$ along $x = z \tan \theta$ for $x >0$, where $ \theta = \sin^{-1}(T)$, so we have a solution for $T  <1$. For $T>0$, the spacetime is the region between the AdS boundary at $z=0$ for $x<0$ and the brane worldvolume $Q$, as pictured in figure \ref{fig:poinc}.\footnote{For $T<0$, the brane has the same position but the spacetime is the wedge between the AdS boundary at $z=0$ for $x>0$ and the brane worldvolume $Q$.} We obtain a useful insight into the interpretation of the tension by considering the entanglement entropy for a region $A$ in the AdS boundary with $x \in (-L,0)$. The bulk RT surface drawn in figure \ref{fig:poinc}, cut off at $z=\epsilon$, gives 
\begin{equation} \label{RTP} 
S(L) = \frac{c}{6} \ln \left(\frac{L}{\epsilon}\right) + \ln \mathfrak g, 
\end{equation}
where $c = \frac{3}{2G}$ is the central charge of the dual two-dimensional CFT, and 
\begin{equation}
\ln \mathfrak g =  \frac{1}{4G} \tanh^{-1} T.   
\end{equation}
The first contribution to the entropy comes from the part of the surface at $x<0$, while the second part comes from the part at $x>0$. The first contribution is the usual entanglement entropy for an interval of length $L$ in a two-dimensional CFT, while the second term is the boundary entropy which appears in BCFTs.\footnote{From the CFT perspective, there is a boundary state $|B \rangle$ associated to the boundary at $x=0$, and $\mathfrak g = \langle 0 | B \rangle$.} Thus, $\ln \mathfrak g$ can be viewed as a boundary analogue of the central charge, and we see that $T$ controls this central charge, analogously to the relation between $\ell$ and $c$. 

It is useful to consider the regime $T \approx 1$, when $\theta \approx \pi/2$, and the end of the world brane can be viewed as a cutoff version of the AdS boundary. This corresponds to a limit where the number of boundary degrees of freedom  is large,  $\ln \mathfrak g \gg c$. We can then obtain an effective gravitational theory on the brane by integrating over the bulk spacetime. In the higher-dimensional context, this gives a Karch-Randall theory, with an effective Einstein action on the brane \cite{Randall:1999ee,Randall:1999vf,Karch:2000ct}. In the present two-dimensional case, this gives a non-local gravity action \cite{Skenderis:1999nb}, as discussed in \cite{Chen:2020uac}. This can be written in a local form by introducing an auxiliary scalar field; the brane gravity action is then a Polyakov action 
\begin{equation} \label{Ipoly} 
I_{Poly} = \frac{1}{32\pi G} \int d^2 y \sqrt{-h} \left[ - \frac{1}{2} h^{ab} \nabla_a \phi \nabla_b \phi + \phi \, {}^{(2)} R - 2 e^{-\phi} \right], 
\end{equation}
where ${}^{(2)} R$ is the Ricci scalar of the metric $h_{ab}$ on the brane.     

In earlier holographic analyses, attention focused on the relation between the bulk spacetime description, where we have Einstein gravity coupled to a constant-tension brane, and the boundary perspective, where we have the CFT dual to the bulk theory with some boundary state $|B\rangle$ at $x=0$. The novelty in recent work, such as in \cite{Almheiri:2019hni}, is to highlight an intermediate effective theory, where we integrate over the bulk spacetime to obtain an effective gravity theory on the brane; we then have a non-gravitational CFT for $x<0$ joined across an interface to the same CFT coupled to the brane gravity theory \eqref{Ipoly} for $x>0$. This is then a useful model to study the appearance of islands in the gravitational theory, by relating them to the classical RT surfaces in the bulk gravity. We can illustrate the idea by relating the entropy \eqref{RTP} to the brane gravity theory. For the Polyakov action \eqref{Ipoly}, the generalized entropy is 
\begin{equation}
    S_{\mathrm{gen}}(\mathcal I) = \frac{\phi_{\partial \mathcal I}}{8G} + S_{\mathrm{eff}}(\mathcal I), 
\end{equation}
where $\phi_{\partial\mathcal{I}}$ is the value of the scalar at the boundary of $\mathcal{I}$ and $\phi_{\partial\mathcal{I}}/2$ is the zero-dimensional analogue of the area. In the solution we considered above, the brane has an AdS$_2$ geometry with $^{(2)} R = - 2/\ell_2^2 = -2(1-T^2)$, and the auxiliary scalar is  
\begin{equation}
    \phi=\ln\left(-\frac{2}{^{(2)} R}\right) = - \ln(1-T^2) \approx 2 \tanh^{-1} T. 
\end{equation}
Thus, we can re-interpret the $\ln \mathfrak g$ term in the entropy \eqref{RTP} as due to the boundary term in brane gravity,
\begin{equation}
    S=\frac{\phi}{8 G}\approx \ln \mathfrak g,
\end{equation}
where this comes from the boundary of the region in the brane that is included as an island for the region we consider in the boundary CFT. 

To investigate gravity in closed universes, \cite{Cooper:2018cmb} considered a solution with a brane in one asymptotic region of an eternal black hole. We consider the simplest version, with  a brane in a BTZ geometry. Considering first the Euclidean solution, we take the bulk metric to be 
\begin{equation} \label{btz} 
ds^2 = (r^2-r_h^2) d \tau^2 + \frac{dr^2}{(r^2-r_h^2)} + r^2 d \phi^2, 
\end{equation}
where $\tau$ is periodic with period $\beta = 2\pi/r_h$ to make the metric regular at $r=r_h$. We consider a brane which respects the $U(1)$ symmetry along $\phi$, so the position of the brane is parametrized by $r(\tau)$. The trajectory of the brane is given by \cite{Cooper:2018cmb}
\begin{equation}
    \sqrt{\frac{r^2}{r_h^2}-1}\cos (r_h\tau)=\frac{T}{\sqrt{1-T^2}}.
\end{equation}
The brane intersects the AdS boundary at $r \to \infty$ at $\tau = \pm \beta/4$, and reaches a minimum radius $r=r_0$ at $\tau=0$, where 
\begin{equation} \label{r0}
 r_0 = \frac{r_h}{\sqrt{1-T^2}}. 
\end{equation}
For $T>0$, the spacetime includes $r=r_h$, so the $\tau=0$ slice includes the whole of one asymptotic region of the black hole, and a portion of the other region, up to $r=r_0$. In the boundary, the intersection of the brane with the asymptotic AdS boundary corresponds to a boundary state $|B\rangle$ for the CFT, and the Euclidean evolution defines a state $| \Psi \rangle = e^{-\beta H/4} |B \rangle$. In the regime where this bulk solution dominates the path integral with these boundary conditions, this state is dual to the bulk $\tau=0$ time slice.

This time slice provides initial conditions for a Lorentzian evolution, which is just obtained by analytically continuing in time. The bulk metric is the Lorentzian BTZ black hole, and  the brane now follows a trajectory given by 
\begin{equation}
    \sqrt{\frac{r^2}{r_h^2}-1}\cosh (r_h t)=\frac{T}{\sqrt{1-T^2}},
\end{equation}
which reaches a maximum radius $r=r_0$ at $t=0$, and falls into the black hole and meets the singularity at $r \to 0$ in the past and future. Thus, the brane worldvolume is a closed universe undergoing a big-bang, big-crunch FRW cosmology. 

Fundamentally, this bulk spacetime is described by the two-dimensional CFT in the state $|\Psi \rangle$, but we can also consider it from the intermediate perspective, where we integrate over the bulk spacetime to obtain an effective theory where we have a CFT coupled to gravity on the brane, entangled with the CFT on the AdS boundary. The appropriate semi-classical state is a deformation of the usual dual description of BTZ, in terms of an entangled thermofield double state \cite{Maldacena:2001kr}, 
\begin{equation} \label{tfd} 
| TFD \rangle = \sum_i e^{-\beta E_i/2} |i\rangle_L |i\rangle_R.
\end{equation}
Considering the situation with an end of the world brane, at least for $T \approx 1$, can be understood as turning on gravity in one of the two copies with the effective action \eqref{Ipoly}. So in the semi-classical picture, we have two copies of the CFT, one in a universe with dynamical gravity and one in a fixed background, in an entangled state which is a deformation of \eqref{tfd}. 

In this semi-classical effective theory, we apply the island rule to conclude that the entanglement in the semi-classical state implies that the whole of the gravitating universe is included in an island for the boundary CFT. From the fundamental microscopic point of view, this is just the statement that the whole spacetime, including the brane system, is encoded in the fundamental state $|\Psi \rangle$ of the dual CFT; fundamentally the brane is not something independent which is entangled with the CFT, but rather is a part of it. There is presumably some code subspace of states in the CFT built on $|\Psi \rangle$ which encodes excitations in the semi-classical Hilbert space in the closed universe on the brane. 

\section{Review of multiboundary wormholes}
\label{wormhole} 

We want to generalize the previous model to a situation where the semi-classical theory has a gravitating brane entangled with several non-gravitating systems. We will build this model by adding an end of the world brane to the multiboundary wormhole solutions discussed in \cite{Brill:1995jv,Aminneborg:1997pz,Krasnov:2000zq,Krasnov:2003ye,Skenderis:2009ju,Balasubramanian:2014hda,Marolf:2015vma}. In this section we  give a brief review of  the multiboundary wormhole solutions in three-dimensional gravity, and their relation to the Euclidean path integral in the dual two-dimensional CFT. 

We consider solutions of the action \eqref{I3d}, so the bulk spacetime is locally AdS$_3$. We can obtain Lorentzian solutions with multiple asymptotic boundaries by considering quotients of AdS$_3$ by some discrete subgroup of its isometry group $SL(2,\mathbb{R}) \times SL(2,\mathbb{R})$. We will describe the quotient using the $SL(2,\mathbb{R})$ representation of AdS$_3$, following \cite{Maxfield:2014kra}.\footnote{For another perspective on the construction of the three-boundary wormhole see \cite{Caceres:2019giy}.}  A point in the spacetime is described by an $SL(2,\mathbb{R})$ matrix $p$, and the spacetime metric is $ds^2 = - \det(dp)$. We can parametrize $p$ in terms of coordinates in an $\mathbb{R}^{2,2}$ embedding space, 
\begin{equation}
   p=\begin{pmatrix}
    X_0+X_2 & X_3-X_1 \\
    X_3+X_1 & X_0-X_2
    \end{pmatrix}.
\end{equation}
This has $\det p=1$ if the embedding space coordinates satisfy $-X_0^2-X_1^2+X_2^2+X_3^2 = -1$. The global coordinates on AdS$_3$ are given by  
\begin{alignat}{2}
    X_0 & = \cosh \chi \cos t, \qquad && X_1 = \cosh \chi \sin t,  \\
    X_2 & = \sinh \chi \cos \theta, \qquad && X_3 = \sinh \chi \sin \theta. 
\end{alignat}
The metric in global coordinates is 
\begin{equation}
    ds^2 = - \cosh^2 \chi dt^2 + d\chi^2 + \sinh^2 \chi d\theta^2. 
\end{equation}
The $SL(2,\mathbb{R})_L \times SL(2, \mathbb{R})_R$ isometries act as  $p \to g_L p g_R^t$. We will focus on the diagonal subgroup acting as $p \to g p g^t$. This maps symmetric $p$ to symmetric $p$, so it leaves the surface at $X_1 =0$, corresponding to $t=0$ in global coordinates, invariant. Thus, a quotient by a discrete subgroup $\Gamma$ of this diagonal $SL(2,\mathbb{R})$ preserves the time-reflection symmetry about this surface, and we can define a Euclidean continuation which is the quotient of $\mathbb{H}^3$ by the same group $\Gamma$. We will largely focus on the action of the quotient on the surface $X_1=0$, which is described in global coordinates as the Poincar\'e disc. 

\begin{figure}[ht]
\begin{subfigure}{.5\textwidth}
  \centering
  \includegraphics[width=.8\linewidth]{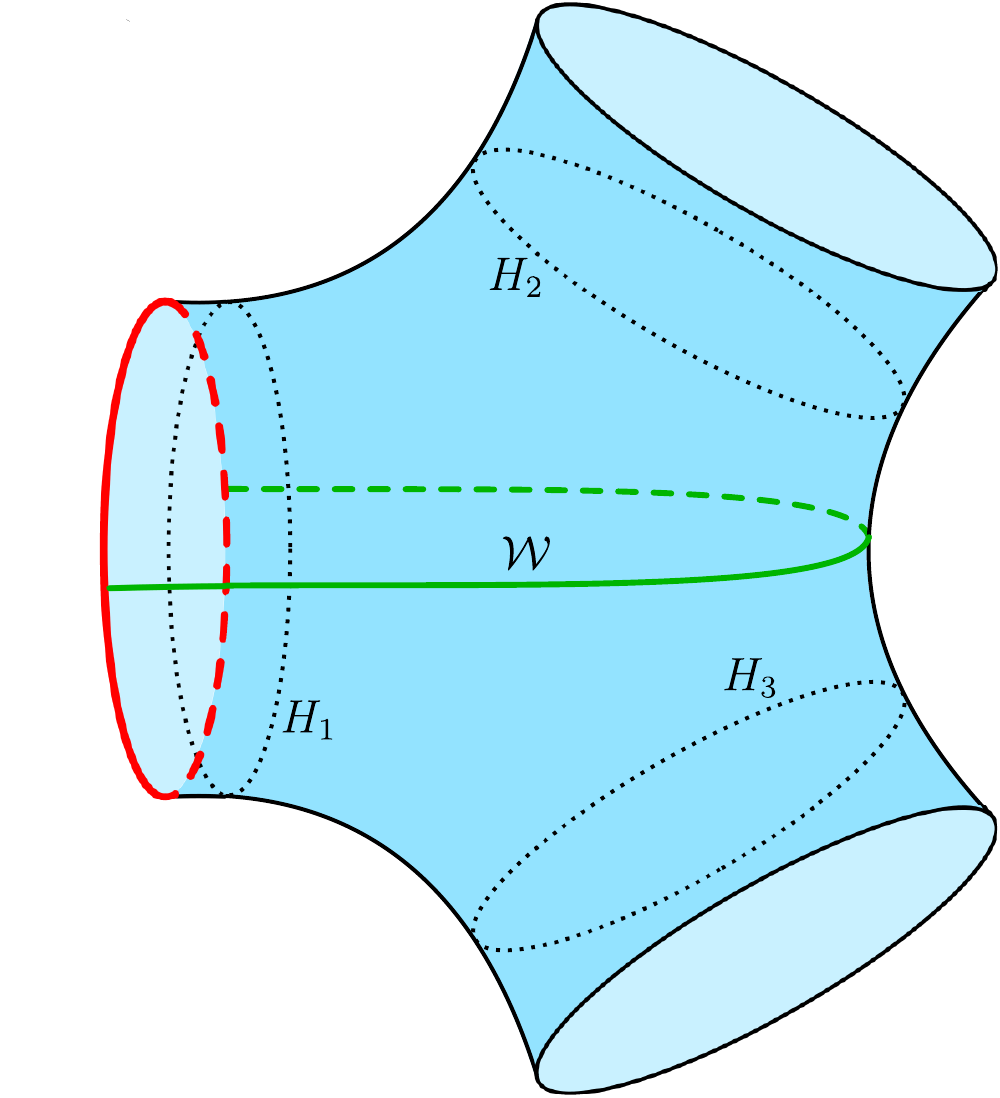}  
  \label{fig:sub-first}
\end{subfigure}
\begin{subfigure}{.5\textwidth}
  \centering
  \includegraphics[width=.8\linewidth]{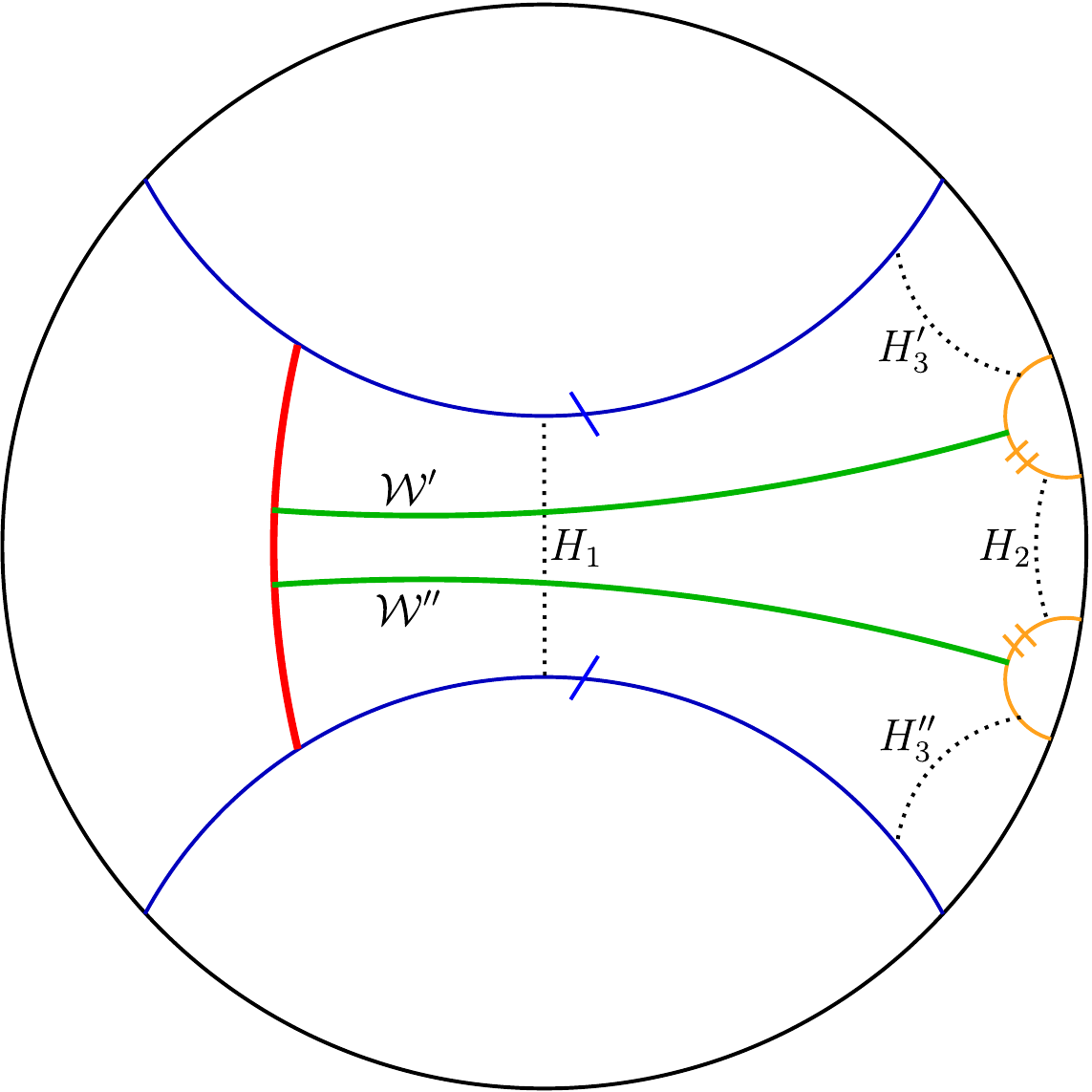}  
  \label{fig:sub-second}
\end{subfigure}
\caption{The pair of pants geometry with an end of the world brane (red) in one asymptotic region. On the left is a cartoon of the geometry of the $t=0$ surface, and on the right is its description as a quotient of the Poincar\'e disc model. In the right picture the central region bounded by the blue and orange geodesics is a fundamental region for the identification. The geodesic $\mathcal{W}\equiv\mathcal{W}'\cup\mathcal{W}''$, which is the minimal geodesic anchored on the end of the world brane running in between the two asymptotic region, is shown in green. The horizon $H_3$ is similarly defined as $H_3\equiv H_3'\cup H_3''$.}
\label{fig:fig}
\end{figure}

We will consider the geometry with three asymptotic boundaries, formed by a quotient by a subgroup $\Gamma$ with two hyperbolic generators. We parametrize these generators as  
\begin{equation}
    g_1=\begin{pmatrix}
        \cosh(\frac{\ell_1}{2}) & \sinh(\frac{\ell_1}{2}) \\
        \sinh(\frac{\ell_1}{2}) & 
        \cosh(\frac{\ell_1}{2})
        \end{pmatrix},
        \quad\quad
    g_2=\begin{pmatrix}
        \cosh(\frac{\ell_2}{2}) & e^{\omega}\sinh(\frac{\ell_2}{2}) \\
        e^{-\omega}\sinh(\frac{\ell_2}{2}) & 
        \cosh(\frac{\ell_2}{2})
        \end{pmatrix}.
\end{equation}
The generator $g_1$ acts as a boost in the $X_0 -X_3$ plane in the embedding space. The generator $g_2$ can be written as a similar boost, up to conjugation, 
\begin{equation}
g_2 = g_\omega \tilde g_2 g_\omega^{-1}, \quad  g_\omega =\begin{pmatrix} e^{\omega/2} & 0 \\ 0 & e^{-\omega/2} \end{pmatrix},\quad \tilde g_2 = \begin{pmatrix}
        \cosh(\frac{\ell_2}{2}) & \sinh(\frac{\ell_2}{2}) \\
        \sinh(\frac{\ell_2}{2}) & 
        \cosh(\frac{\ell_2}{2})
        \end{pmatrix}. \quad 
\end{equation}
The conjugating matrix $g_\omega$ acts as a boost in the $X_0 - X_2$ plane. If we considered just the quotient by the abelian group generated by $g_1$, the resulting spacetime would be the BTZ black hole \eqref{btz} with $r_h = \ell_1/2\pi$. The BTZ coordinates are related to the embedding coordinates by 
\begin{alignat}{2}
    X_0 & = \frac{r}{r_h}\cosh{(r_h\phi)}, \qquad && X_1 = \sqrt{\frac{r^2}{r_h^2}-1}\sinh{(r_ht)}  \\
    X_3 & = \frac{r}{r_h}\sinh{(r_h\phi)}, \qquad && X_2 = \pm \sqrt{\frac{r^2}{r_h^2}-1}\cosh{(r_ht)}.
\end{alignat}
The quotient by $g_1$ acts as translation in $\phi$, $\phi \to \phi + 2\pi$. In the Poincar\'e disc representation, on the right in figure \ref{fig:fig}, a symmetric fundamental region for this identification is the region between the two blue geodesics, corresponding to $\phi = \pm \pi$. The horizon is the closed geodesic at $X_2=0$, of proper length $\ell_1$. The other generator $g_2$ similarly identifies the two orange geodesics on the right of the picture, and the dotted curve connecting them is a closed geodesic of proper length $\ell_2$, which is a horizon for the asymptotic region on the right. We restrict to $e^\omega > \coth\left( \frac{\ell_1}{4} \right) \coth \left( \frac{\ell_2}{4} \right)$, so that the orange and blue geodesics don't intersect \cite{Maxfield:2014kra}. The region bounded by these geodesics is a fundamental region for the quotient by the group $\Gamma$ generated by $g_1, g_2$.  The quotient geometry has three asymptotic regions, each of which is isomorphic to the BTZ black hole outside a horizon. The horizon in the third asymptotic region is formed of the two minimal geodesics connecting the blue and orange surfaces in the right picture. The length $\ell_3$ of this horizon is given by
\begin{equation} \label{L3}
    \cosh\frac{\ell_3}{2}=\frac{1}{2}\Tr g_3=-\frac{1}{2}\Tr g_1g_2^{-1}=\cosh\omega\sinh\frac{\ell_1}{2}\sinh\frac{\ell_2}{2}-\cosh\frac{\ell_1}{2}\cosh\frac{\ell_2}{2}. 
\end{equation}
The region in between the three horizons is referred to as the causal shadow region, as no causal influence can propagate from this region to the asymptotic regions.  The quotient of the surface $X_1=0$ is thus a surface $\Sigma$ with the topology of a pair of pants,  labelled by three parameters $\ell_1, \ell_2, \ell_3$, as pictured on the left in figure \ref{fig:fig}. This is the simplest example of a multiboundary wormhole. 

We can analytically continue the Lorentzian solution to a Euclidean spacetime with geometry 
\begin{equation} \label{eworm}
    ds^2 = d\rho^2 + \cosh^2 \rho\, d\Sigma^2. 
\end{equation}
The conformal boundary of the Euclidean solution is two copies of the surface $\Sigma$. We can take the Euclidean path integral in the CFT on one copy of $\Sigma$; this defines an entangled state for the CFT in  three copies of the Hilbert space of the CFT on $S^1$,  $|\Sigma \rangle \in \mathcal H_1 \times \mathcal H_2 \times \mathcal H_3$. The bulk geometry provides a candidate bulk saddle-point dual to this Euclidean path integral; when this is the dominant bulk saddle-point, the bulk geometry on $\Sigma$ is the dual of the CFT state $|\Sigma \rangle$. In the state $|\Sigma \rangle$, the reduced density matrix $\rho_i$ on each copy of the CFT is not  thermal, but one-point functions of local operators are determined by the geometry in the exterior region, so they take thermal values. In particular, the stress tensor of the CFT degrees of freedom will be a perfect fluid with coarse-grained entropy $S^{(c)}_i = \frac{\ell_i}{4G}$. The fine-grained entropy of $\rho_i$ is $S_i = \frac{1}{4G} \mbox{min} (\ell_i, \ell_j + \ell_k)$, so if one of the horizons is longer than the sum of the other two, the actual entropy of the density matrix in that region is less than the coarse-grained thermal value. 

\begin{figure}[ht]
\begin{subfigure}{.5\textwidth}
  \centering
  \includegraphics[width=.4\linewidth]{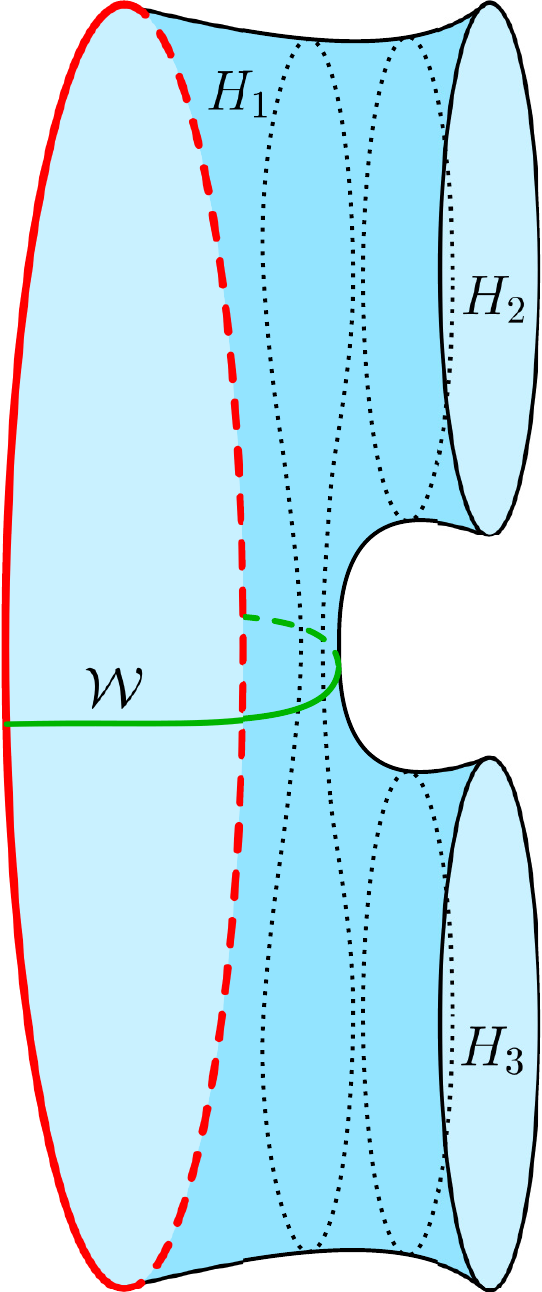}  
  \label{fig:sub-first}
\end{subfigure}
\begin{subfigure}{.5\textwidth}
  \centering
  \includegraphics[width=.8\linewidth]{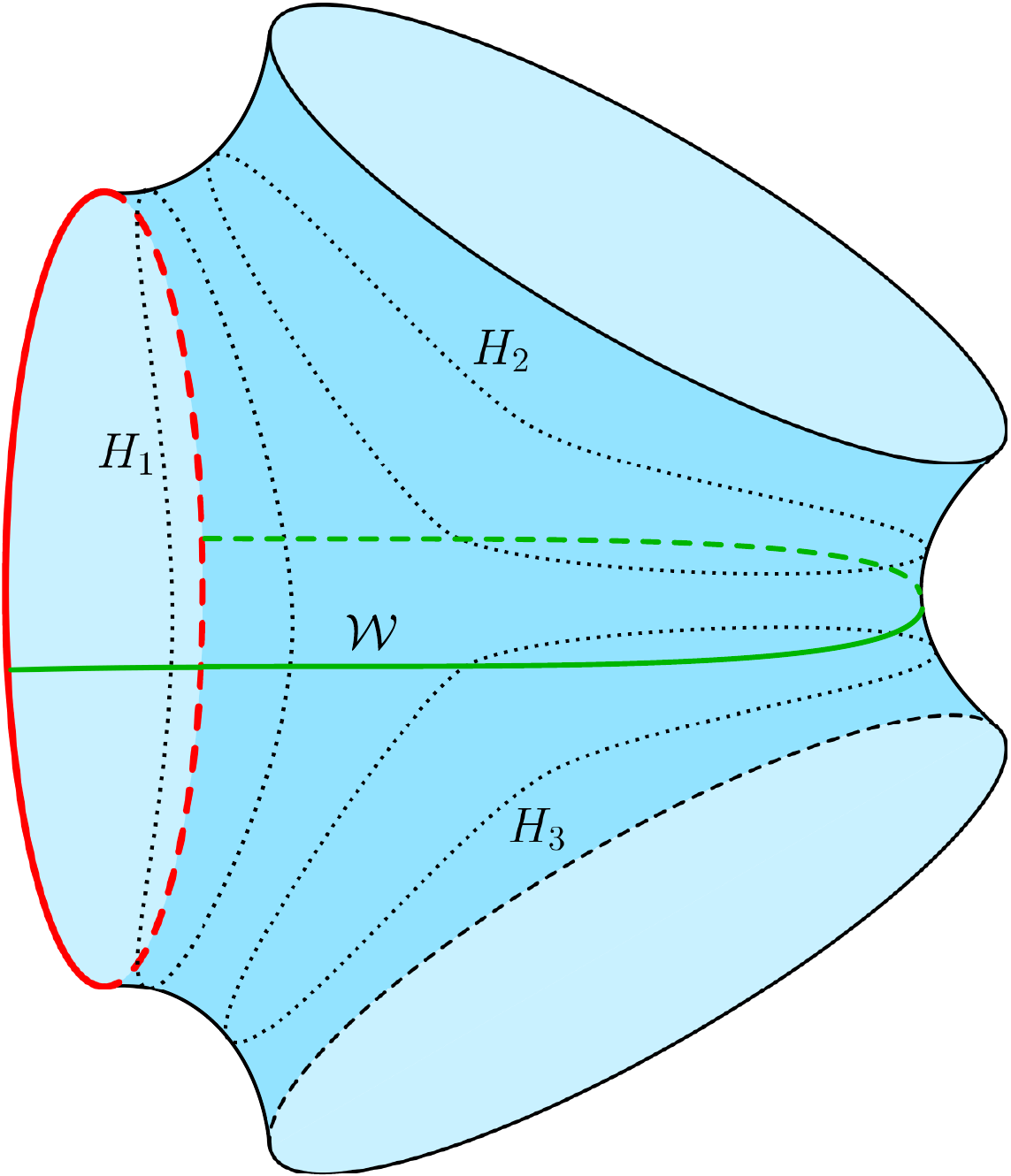}  
  \label{fig:sub-second}
\end{subfigure}
\caption{In the large $\ell_i$ limit, the horizons are locally identified. There are two cases: if one length is larger than the sum of the other two, $\ell_i > \ell_j + \ell_k$, we have an ``eyeglass'' picture, where the whole of the two short horizons are identified with the long one, and the remaining parts of the long horizon are identified with each other. Otherwise, each horizon has a portion which is identified with each of the others.}
\label{fig:largel}
\end{figure}

We will be particularly interested in the limit of large $\ell_1, \ell_2, \ell_3$, with fixed ratios. In \cite{Marolf:2015vma}, the entanglement structure of $|\Sigma \rangle$ was shown to simplify in this limit. The essential point is that the causal shadow region has a constant negative curvature geometry bounded by geodesics, so its area is fixed by the Gauss-Bonnet theorem. Hence, as the horizons become long, the distance between them must become small. We can decompose the path integral over $\Sigma$ defining the state $|\Sigma \rangle$ into an integral over the regions $E_a$ outside the horizons, which are conformal to round cylinders, and the integral over the causal shadow region. The path integral over the causal shadow region then identifies the horizons locally. There are two different regimes, as pictured in figure \ref{fig:largel}: if one of the horizons is longer than the sum of the other two, we have an ``eyeglass'' picture, where the whole of the two short horizons are identified with the long one, and the remaining parts of the long horizon are identified with each other. Otherwise, each horizon has a portion which is identified with each of the others. In the ``eyeglass'' regime, to leading order in $c$ the state $|\Sigma\rangle$ only involves entanglement between the boundaries associated to the short horizons and the long one; there is no entanglement between the two boundaries associated to the short horizons. 

\section{Islands on the braneworld} 
\label{islands} 

We now turn to the main point of our paper: to add a brane to the multiboundary wormhole spacetime, to obtain a model where we can investigate the description of a closed universe entangled with a quantum system with two components. We simply consider inserting the same kind of brane considered above in one of the exterior BTZ regions in the $t=0$ slice, as pictured in figure \ref{fig:fig}, at the maximum radius $r_0$ given by \eqref{r0}. We will call the asymptotic region that is cut off in this way region 1.  This initial data has a Euclidean continuation where the brane intersects the conformal boundary of the Euclidean wormhole spacetime \eqref{eworm}. The brane intersects the Euclidean boundary at $\tau = \pm \beta/4$ in the BTZ coordinates. The path integral over half the Euclidean boundary, with the boundary state $|B \rangle$ dual to the end of the world brane inserted at $\tau = -\beta/4$, defines a state $|\Psi \rangle \in \mathcal H_2 \times \mathcal H_3$ dual to this bulk geometry.

At least when $r_0 \gg r_h$, this geometry can also be described in a semi-classical effective theory where we integrate out the bulk spacetime to obtain a CFT coupled to gravity living on the end of the world brane, entangled with the two other copies of the CFT. In this description, the semi-classical state of the theory is roughly the state $|\Sigma \rangle$ dual to the multiboundary wormhole, but with some deformation from coupling the CFT in boundary 1 to gravity; we will call this deformed state $|\tilde \Sigma \rangle$. The brane is embedded in a BTZ geometry, with horizon length $\ell_1$. The effective stress tensor of the semi-classical theory on the brane is thus a perfect fluid, with a coarse-grained entropy $S_{coarse} = \frac{\ell_1}{4G}$.  

In the state $|\tilde \Sigma \rangle$, we have some entanglement between the degrees of freedom in the closed universe and the two CFTs on boundaries 2 and 3. This entanglement is most easily characterised in the large $\ell_i$ limit, where the entanglement is approximately local; small regions in the brane form a thermofield double state with corresponding regions in one of the other systems (or with another region on the brane). We will explore the  entanglement structure of $|\tilde \Sigma \rangle$ by considering RT surfaces in the bulk spacetime in figure \ref{fig:fig}. Before carrying out a quantitative analysis, we give a general qualitative description. 

Since the microscopic state $|\Psi \rangle \in \mathcal H_2 \times \mathcal H_3$ is pure, the entropy of $\rho_2 = \mbox{Tr}_3 (|\Psi \rangle \langle \Psi|)$ is equal to the entropy of $\rho_3 = \mbox{Tr}_2 (|\Psi \rangle \langle \Psi|)$. There are three candidate RT surfaces for this entropy: the horizon $H_2$, of length $\ell_2$, the horizon $H_3$, of length $\ell_3$, or a geodesic with end points on the end of the world brane which separates the two boundaries, as pictured in figure \ref{fig:fig}. We call the minimal-length geodesic in this class $\mathcal{W}$.  We take without loss of generality $\ell_2 \leq \ell_3$; then the RT surface is either $H_2$ or $\mathcal{W}$. 

When the RT surface is $\mathcal{W}$, the brane is partially encoded in $\mathcal H_2$ and partially in $\mathcal H_3$. This is the situation where the entanglement of the brane with boundaries 2 and 3 in $|\tilde \Sigma \rangle$ is big enough to overcome the cost of having a non-trivial island. In this situation the entropy $S = \frac{\ell_{\mathcal W}}{4G}$ is partially due to the geometric entropy on the brane, and partially due to entanglement between the degrees of freedom in system 2 together with the associated island on the brane and the rest.

When the RT surface is $H_2$, the brane is entirely encoded in $\mathcal H_3$. This is the regime which we are particularly interested in. If we treat $\mathcal H_2$ as a reference system, and trace over it,  semi-classical entanglement between the brane and $\mathcal H_2$ gives us a mixed state on the brane, which is encoded microscopically in the mixed state $\rho_3$ of the CFT on the boundary.  The fine-grained entropy is $S = \frac{\ell_2}{4G}$. Semi-classically, this can include both contributions from entanglement between the brane and boundary 2, and between boundary 3 and boundary 2. In the large $\ell_i$ limit, we can cleanly separate these two contributions, because of the local structure of the entanglement: the portion of $H_2$ which lies along $H_1$ represents entanglement between the brane and boundary 2 in the semi-classical state, so we identify the length of this portion with the entropy $S_{brane}$ of the semi-classical state of the matter on the brane. There are three cases: 
\begin{itemize}
    \item If $\ell_1 > \ell_2 + \ell_3$, in $|\tilde \Sigma \rangle$ boundary 2 is entirely entangled with the brane, with no entanglement with boundary 3. The whole entropy $S_{brane} = S=\frac{\ell_2}{4G}$ can be thought of as entropy of the mixed state on the brane. 
    \item If $\ell_3 > \ell_1 + \ell_2$, boundary 2 is entirely entangled with boundary 3, with no entanglement with the brane. The effective semi-classical state on the brane remains pure on tracing out boundary 2, as it's solely entangled with boundary 3. 
    \item Otherwise, the entanglement is partially with the brane and partially with boundary 3. The entropy of the mixed state on the brane is determined by the portion of horizon 2 that lies along horizon 1, which gives  $S_{brane} = \frac{1}{8G} (\ell_1  +\ell_2 - \ell_3)$. 
\end{itemize}
We see that $S_{brane} \leq S_{coarse}/2$, with equality when $\ell_2=\ell_3 > \ell_1/2$. 

\subsection{The $\mathcal W$ geodesic} 

We now compute the length of the $\mathcal{W}$ geodesic, to be compare it to the horizon $H_2$. By time reflection symmetry the geodesic lies in the $t=0$ slice of the geometry. In the Poincar\'e disc model $\mathcal{W}$ corresponds to two geodesics $\mathcal{W}'$ and $\mathcal{W}''$, which hit the orange identification surfaces. There is a $\mathbb{Z}_2$ reflection symmetry about the $\theta = 0,\,\pi$ axis in the Poincar\'e disc, corresponding to the reflection symmetry about the plane of the page in the pair of pants geometry shown in figure \ref{fig:fig}. This $\mathbb{Z}_2$ symmetry implies that the two geodesics are identical and meet the identification surfaces orthogonally, so as to produce a smooth connected curve in the quotient space. The end of the world brane sits at constant $r$, and the geometry between the brane and the identification surface is locally BTZ, so if we consider connecting a fixed point on the identification surface to the end of the world brane, the the length is minimal when the curve lies at constant $\phi$. The $\mathcal{W}'$ geodesic is then obtained by varying the point on the identification surface, to find the curve of constant $\phi$ which meets the identification surface orthogonally.
The identification surfaces are geodesics in the Poincar\'e disc, given by  
\begin{equation}\label{geod}
    \tanh\chi \cos(\theta-\alpha)=\cos\psi,
\end{equation}
in global coordinates. By finding the point where the normal to the identification surface $n \propto dr$, we find the angle $\theta_1$ at which the geodesic hits the identification surface. It will be more convenient to define $\vartheta\equiv\theta-\alpha$. We get
\begin{equation}
    \tan\vartheta_1=\tan\alpha\tan^2\psi.
\end{equation}
The geodesic distance between two spacelike separated points $s(X,X')$ is
\begin{equation}
    \cosh s(X,X')=-X\cdot X'.
\end{equation}
The length of $\mathcal{W}$ is then given by
\begin{equation}
 \cosh\frac{1}{2}\ell_{\mathcal{W}} =\frac{\csc\psi}{\sqrt{1-T^2}} \left(\cos \alpha +\frac{T}{2} \sqrt{\cos 2 \alpha +\cos 2 \psi }\right). 
\end{equation}

To relate this to the horizon lengths, we need to relate the parameters $\alpha, \psi$ specifying the geodesics identified by $g_2$ to the horizon lengths. The geodesics identified by $g_1$ have $\tanh \chi \cos(\theta\pm\pi/2) = \cos \delta$, where 
\begin{equation}
\cos \delta = \tanh \chi_{min} = \tanh \frac{\ell_1}{2} \Rightarrow \sin \delta = \sech \frac{\ell_1}{2}. 
\end{equation}
The geodesics identified by $g_2$ are similar, but conjugated by $g_\omega$. That is, there is a conjugated Poincar\'e coordinate system with $\tilde{\alpha}=\pi/2$, $(\sin\tilde{\psi})^{-1}=\cosh\frac{\ell_2}{2}$. Conjugation by $g_\omega$ acts on the boundary coordinates by 
\begin{equation}
    \begin{pmatrix} \cos \frac{\theta}{2} \\ \sin \frac{\theta}{2} \end{pmatrix} \propto g_\omega \begin{pmatrix} \cos \frac{\tilde{\theta}}{2} \\ \sin \frac{\tilde{\theta}}{2} \end{pmatrix}.
\end{equation}
This gives 
\begin{equation} \label{a/p}
    \sin\alpha=\frac{1}{\sqrt{\cosh^2\omega-\sinh^2\omega\sin^2\tilde{\psi}}}, \quad\quad \sin\psi=\sin\tilde{\psi}\sin\alpha.
\end{equation}
Thus,
\begin{equation} \label{ell2}
    \cosh\frac{\ell_2}{2}=\frac{1}{\sin\tilde{\psi}}=\frac{\sin\alpha}{\sin\psi}.
\end{equation}
Using \eqref{L3}, we find 
\begin{equation}
    \tan\alpha=\frac{\sqrt{2}\cosh\frac{\ell_2}{2}\sinh\frac{\ell_1}{2}}{\sqrt{1+\cosh\ell_1+\cosh\ell_2+\cosh\ell_3+4\cosh\frac{\ell_1}{2}\cosh\frac{\ell_2}{2}\cosh\frac{\ell_3}{2}}}.
\end{equation}

This gives an expression for the length of the $\mathcal W$ geodesic as a function of the horizon lengths and the tension $T$,
\begin{align} \nonumber
  \cosh\frac{1}{2}\ell_{\mathcal{W}}= \frac{\csch\frac{\ell_1}{2}}{\sqrt{2(1-T^2)}}&\left(\sqrt{1+\cosh\ell_1+\cosh\ell_2+\cosh\ell_3+4\cosh\frac{\ell_1}{2}\cosh\frac{\ell_2}{2}\cosh\frac{\ell_3}{2}}\right. \\
  & \quad +T\left.\sqrt{2+\cosh\ell_2+\cosh\ell_3+4\cosh\frac{\ell_1}{2}\cosh\frac{\ell_2}{2}\cosh\frac{\ell_3}{2}}\right).
\end{align}
We can now compare this to the horizon lengths. Since we assume $\ell_2 \leq \ell_3$, we want to know whether $\ell_{\mathcal W}$ is bigger or smaller than $\ell_2$. 

\begin{figure}[h]
    \centering
    \includegraphics[scale=0.6]{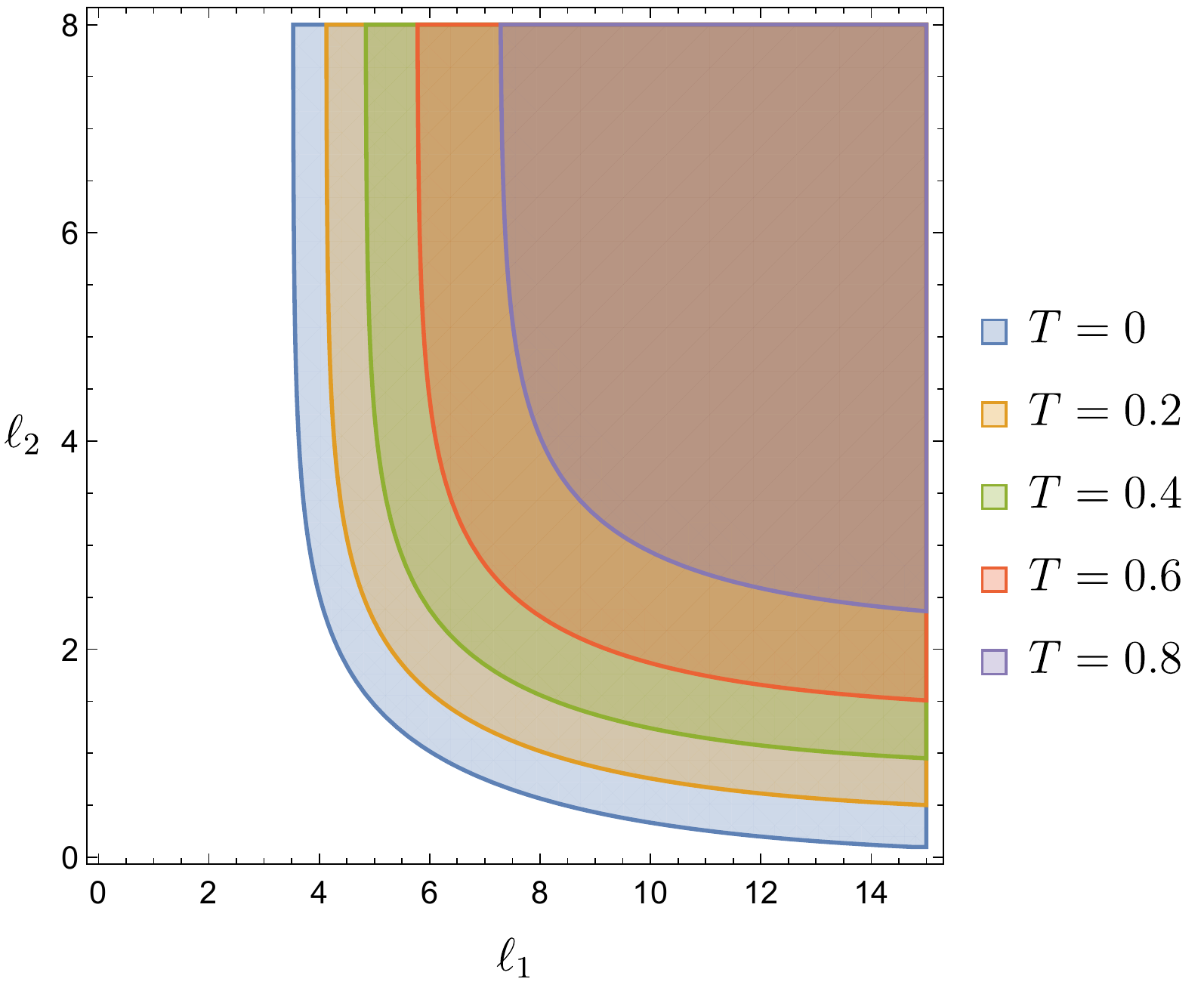}
    \caption{The regions where $\ell_{\mathcal W} < \ell_2$ for $\ell_2=\ell_3$ for various values of $T$}
    \label{fig:my_label}
\end{figure}

The length $\ell_{\mathcal{W}}$ is monotonically decreasing in $\ell_1$, and monotonically increasing in $\ell_2, \ell_3$. The former gives us a lower bound for the length: taking the large $\ell_1$ limit at fixed $\ell_2, \ell_3$, $ \lim_{\ell_1\to\infty} \ell_{\mathcal{W}}=2\tanh^{-1}T$, so $\ell_{\mathcal{W}} \geq 2 \tanh^{-1} T$ for any values of the $\ell_i$. Thus,
\begin{equation}
   \ell_2 < 2 \tanh^{-1} T \quad \Rightarrow \quad \ell_{\mathcal W} > \ell_2;
\end{equation}
for sufficiently small $\ell_2$ the minimal geodesics is always $H_2$. Since $\ell_{\mathcal W}$ is monotonically increasing in $\ell_3$, the minimum value for fixed $\ell_2$ is at $\ell_2 = \ell_3$, where the formula simplifies to
\begin{equation} \label{min}
  \cosh\frac{1}{2}\ell_{\mathcal{W}}=  \frac{\csch{\frac{\ell_1}{4}}}{\sqrt{1-T^2}}\left( \sqrt{\cosh^2\frac{\ell_2}{2} +\sinh^2{\frac{\ell_1}{4}}}+T\cosh{\frac{\ell_2}{2}} \right).
\end{equation}
Here it is interesting to note that 
\begin{equation}
    \ell_1 < 4\sinh^{-1}\left(\sqrt{\frac{1+T}{1-T}}\right) \quad \Rightarrow \quad \ell_{\mathcal W} > \ell_2, 
\end{equation}
so also for sufficiently small $\ell_1$ the minimal geodesic is always $H_2$. In figure \ref{fig:my_label}, we plot the regions where $\ell_{\mathcal W} < \ell_2$ for $\ell_2=\ell_3$ for various values of $T$. 

We can simplify the expression by working in a regime where $\ell_{\mathcal W}$ is large --- this is achieved either by considering large values of the $\ell_i$, or $T \to 1$, or both. Let us first consider the limit $T \to 1$, for general $\ell_i$. This corresponds to taking the brane towards the asymptotic boundary in region 1, so the $\mathcal W$ geodesic gets longer as it extends further out into this region. In this limit we can write
\begin{equation}
    \ell_{\mathcal W} \approx \ell_S + \ln \frac{2}{1-T} \approx \ell_S + 2 \tanh^{-1} T, 
\end{equation}
where 
\begin{align} \nonumber
  e^{\frac{\ell_S}{2}} = \frac{\csch\frac{\ell_1}{2}}{\sqrt{2}}&\left(\sqrt{1+\cosh\ell_1+\cosh\ell_2+\cosh\ell_3+4\cosh\frac{\ell_1}{2}\cosh\frac{\ell_2}{2}\cosh\frac{\ell_3}{2}}\right. \\
  & \quad +\left.\sqrt{2+\cosh\ell_2+\cosh\ell_3+4\cosh\frac{\ell_1}{2}\cosh\frac{\ell_2}{2}\cosh\frac{\ell_3}{2}}\right).
\end{align}
The entanglement entropy when the $\mathcal W$ geodesic is minimal is then 
\begin{equation}
    S_\mathcal{W}=\frac{\ell_\mathcal{W}}{4 G}=\frac{\ell_S}{4 G}+2\ln \mathfrak{g},
\end{equation}
where $\ln \mathfrak{g}=\tanh^{-1}(T)/4G$ is the boundary entropy of the end of the world brane. As in the simpler Poincar\'e-AdS solution discussed in section \ref{braneworld}, the $2 \ln \mathfrak g$ contribution can be seen from the brane gravity perspective as the boundary contribution to the generalized entropy from the boundaries of the island on the brane (there is a factor of 2 here compared to the discussion in section \ref{braneworld} because the island here has two boundaries). 

If we also consider the limit of large horizon lengths, $\ell_{1,2,3} \gg 1$, then $\ell_S$ will typically also be large. In the large horizon length limit, $\ell_S$ can be interpreted as the portion of the $\mathcal W$ geodesic in the ``shadow region'' beyond the horizon, running over the pair of pants between $H_2$ and $H_3$, while the $\ln \mathfrak g$ term comes from the portion of the geodesic between the brane and $H_1$. 

There are three distinct cases in the large horizon length limit: 
\begin{itemize}
\item If $\ell_1 > \ell_2 + \ell_3$, the dominant contribution to $\ell_{\mathcal W}$ comes from the $\cosh \ell_1$ in the first square root; then 
\begin{equation} 
  \cosh\frac{1}{2}\ell_{\mathcal{W}}\approx \frac{1}{ \sqrt{1-T^2}},
\end{equation}
or 
\begin{equation} 
  \ell_{\mathcal{W}} \approx 2\cosh^{-1}\left(\frac{1}{ \sqrt{1-T^2}}\right) = 2 \tanh^{-1} T. 
\end{equation}
That is, in this case $\ell_S$ is small, as part of $H_1$ lies along another portion of $H_1$, so the minimal geodesic over the pair of pants is short. In this regime the $\mathcal W$ geodesic is minimal as soon as $\ell_2 > 2 \tanh^{-1} T$. 
\item If $\ell_3 > \ell_1 + \ell_2$, the dominant contribution to $\ell_{\mathcal W}$ comes from the $\cosh \ell_3$ in both square roots; then 
\begin{equation}  \label{eye1} 
  \cosh\frac{1}{2}\ell_{\mathcal{W}}\approx e^{\frac{\ell_3-\ell_1}{2}} \frac{\sqrt{1+T}}{ \sqrt{1-T}},
\end{equation}
that is 
\begin{equation} \label{eye2}
 \ell_{\mathcal{W}}\approx \ell_3 - \ell_1 + 2 \ln \left( \frac{2\sqrt{1+T}}{ \sqrt{1-T}} \right) \approx \ell_3 - \ell_1 + 2 \tanh^{-1} T, 
\end{equation}
where we again drop an order one term. That is, $\ell_S = \ell_3 - \ell_1$: the $\mathcal W$ geodesic follows the part of $H_3$ that is not along $H_1$. In this regime the $\mathcal W$ geodesic is never minimal, as the first term on the RHS is already bigger than $\ell_2$ by assumption. 
\item Otherwise, the dominant contribution to $\ell_{\mathcal W}$ comes from the final term in both square roots; then 
\begin{equation} 
  \cosh\frac{1}{2}\ell_{\mathcal{W}}\approx e^{\frac{\ell_3-\ell_1-\ell_2}{4}} \frac{\sqrt{1+T}}{ \sqrt{1-T}},
\end{equation}
that is 
\begin{equation} \label{gen}
 \ell_{\mathcal{W}}\approx \frac{1}{2} (\ell_3 + \ell_2 -\ell_1) + 2 \ln \left( \frac{2\sqrt{1+T}}{ \sqrt{1-T}} \right) \approx \frac{1}{2} (\ell_3 + \ell_2 -\ell_1)+ 2 \tanh^{-1} T, 
\end{equation}
again dropping order one terms. That is, $\ell_S = \frac{1}{2} (\ell_3 + \ell_2 -\ell_1)$. The $\mathcal W$ geodesic lies along the section where $H_2$ and $H_3$ lie along each other, as pictured in figure \ref{fig:largel}.  In this regime the $\mathcal W$ geodesic is minimal if 
\begin{equation} 
\ell_2 > \ell_3 - \ell_1 + 4 \tanh^{-1} T. 
\end{equation}
Note that we have not assumed that $\ell_3 - \ell_1$ is positive, but we have assumed $\ell_3 + \ell_2 - \ell_1$ is positive, so this is a stronger condition than the general condition $\ell_2 > 2 \tanh^{-1} T$. 
\end{itemize}

We now have a full picture of the behaviour of the entanglement structure as a function of $\ell_i, T$, at least in the region where all geodesics are long. To illustrate this, let's consider what happens as we change $\ell_2$ at fixed $\ell_1, \ell_3, T$. There are two different cases: $\ell_3 > \ell_1$ and $\ell_1 > \ell_3$. 

If $\ell_1 > \ell_3$, at small $\ell_2$ some of the degrees of freedom on the brane are actually entangled with other regions on the brane, as we are in an ``eyeglass'' situation. The reference system boundary 2 is entirely entangled with the brane in the semi-classical state, and tracing over it gives us a mixed state on the brane of entropy $S_{brane} = \frac{\ell_2}{4G}$. This gives a model of the encoding of a mixed state in a closed universe. As we increase $\ell_2$, we will eventually make a transition to $\mathcal W$ being minimal, and there is an island on the brane associated to boundary 2. If $\ell_1 - \ell_3 > 2 \tanh^{-1} T$, we make the transition at $\ell_2 = 2 \tanh^{-1} T$, before we get out of the ``eyeglass'' situation; otherwise, we first enter a regime where boundary 2 is partially entangled with boundary 3, and the entropy due to entanglement with the brane is $S_{brane} = \frac{1}{8G} (\ell_1  +\ell_2 - \ell_3)$. If $\ell_1 > 4\tanh^{-1} T$, we make a transition to $\mathcal W$ being minimal at $\ell_2 = \ell_3 - \ell_1 + 4 \tanh^{-1} T$. Otherwise, we reach $\ell_2 = \ell_3$ and make a transition to $H_3$ being minimal, and the brane gets entirely encoded in boundary 2.

If $\ell_3 > \ell_1$, at small $\ell_2$ we are in an ``eyeglass'' situation where horizon 3 is long, so initially boundary 2 is entangled only with boundary 3, and the effective semi-classical state on the brane is entangled only with boundary 3. Microscopically, $\mathcal H_3$ factors into a piece which encodes the brane state and a piece which carries the entanglement with boundary 2. $H_2$ remains the minimal geodesic until we reach a regime with non-zero $S_{brane} = \frac{1}{8G} (\ell_1  +\ell_2 - \ell_3)$, and the transition to $\mathcal W$ being minimal is again at $\ell_2 = \ell_3 - \ell_1 + 4 \tanh^{-1} T$ for $\ell_1 > 4\tanh^{-1} T$. 

In all cases, when there is a transition to $\mathcal W$ being minimal the transition is at $S_{brane} = \frac{1}{2G} \tanh^{-1} T = 2 \ln \mathfrak g$. Thus, we learn that the entropy of the effective state on the brane is bounded by $S_{brane} < \frac{1}{2} S_{coarse}$ (otherwise it would be favourable to have the brane entirely encoded in the reference system rather than boundary 3) and $S_{brane} < 2 \ln \mathfrak g$ (otherwise it would be favourable to have an island on the brane). The simplicity of the latter bound is due to our model describing a particular kind of mixed state, where the reference system is entangled with a particular local region on the brane. We could certainly imagine entangling the brane with the reference system in more complicated ways, which could relax this bound. The first bound seems more universal. 

\section{Discussion} 
\label{disc}

We have shown that entangled or mixed states of closed universes can be described in the context of the island formula, if we consider a closed universe entangled with a non-gravitating system with multiple components. We can have situations where the closed universe is encoded in one system, but still has some entanglement with another, which corresponds microscopically to a part of the entanglement between the two systems. We studied a simple model in three dimensions, based on multiboundary wormholes, and found that the portion of the entropy that could be assigned to the state on the brane was bounded by half of the coarse-grained entropy of the effective theory on the brane, and also bounded by the gravitational entropy of an island on the brane. 

The relation to the gravitational entropy on the brane is interesting, and deserves further exploration; it would be interesting if this could shed some light on the interpretation of horizon entropy in closed universes. However, the simple relationship obtained in our model depends on the assumption that the degrees of freedom on the brane that are entangled with the CFT are localised in a particular region of the brane. We would expect that the semi-classical theory would include more general mixed states on the brane with less localised entanglement, by considering more abstract reference systems or systems with more components. 

Another interesting way to use this model is to take $\ell_2$ and $\ell_3$ much smaller than $\ell_1$; the brane then has a large coarse-grained entropy, but the fundamental microscopic description has a much smaller entropy, reminiscent of a `bag-of-gold' spacetime \cite{Fu:2019oyc}. This could be an interesting avenue to explore the reconstruction of states on the brane, using the Petz map \cite{Chen:2019gbt,Penington:2019kki} or the tensor network ideas of \cite{Almheiri:2018xdw}. It would also be interesting to explore the relation of the model we used here to the somewhat different model in \cite{Balasubramanian:2020hfs}. Finally, if we take all the horizon lengths small, the system should also have a non-trivial phase structure; it would also be interesting to explore the description of the brane universe in cases where it's not connected to the conformal boundaries in the bulk spacetime. 

\begin{figure}[ht]
\centering
\includegraphics[scale=0.7]{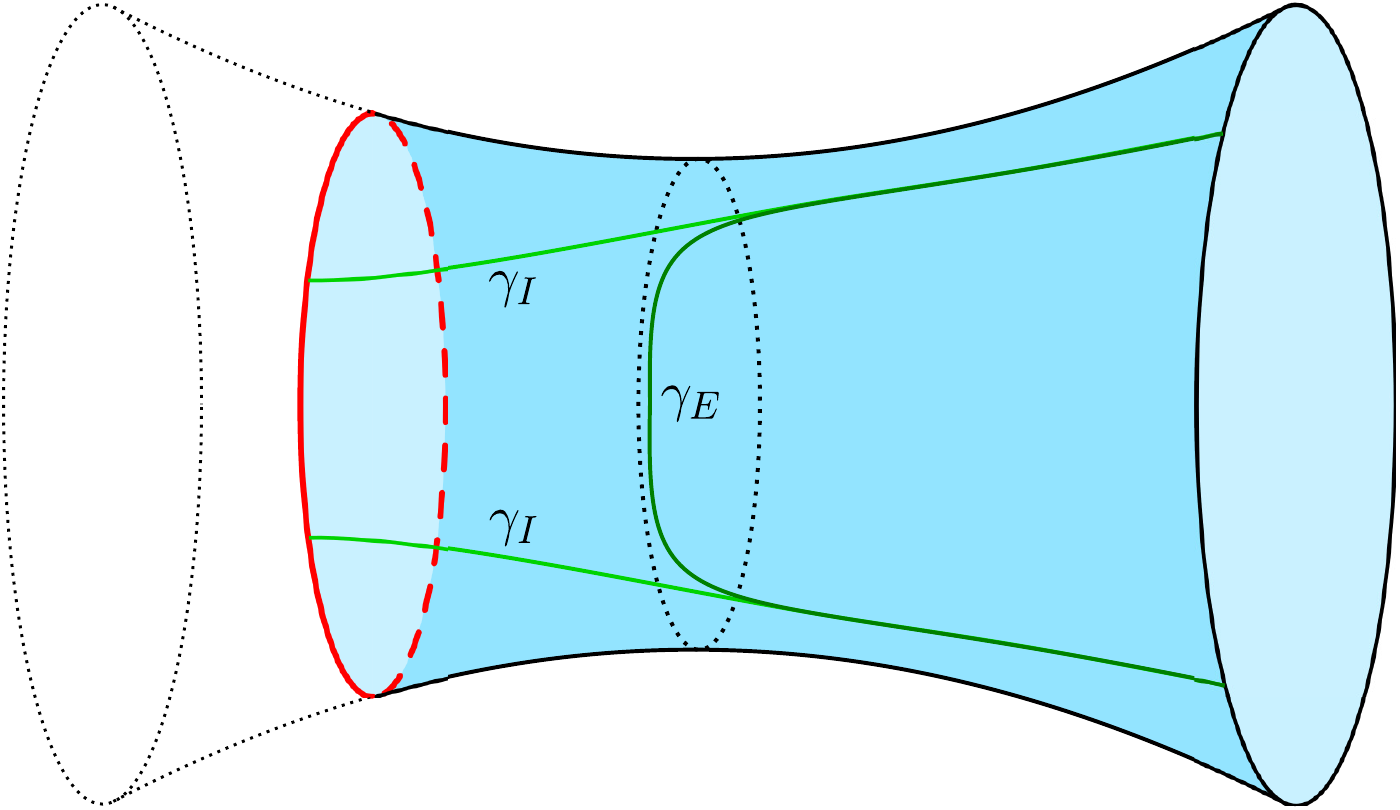}
\caption{For a brane in an eternal black hole spacetime, when we consider a subregion of the CFT on the boundary, the RT surface is either outside the horizon or crosses the horizon and ends on the brane. For large black holes, the surface outside the horizon has a portion which lies along the horizon, whose area can be interpreted as entropy of the mixed state on the brane.}
\label{fig:highd}
\end{figure}

It would be interesting to carry out similar calculations in higher dimensions. There is no simple higher-dimensional version of the multiboundary wormholes, but we could consider working in the original model of \cite{Cooper:2018cmb}, where we introduce a brane in an eternal black hole spacetime. The brane is then encoded in the CFT on the asymptotic AdS boundary, and we can consider dividing this boundary into two regions. There are similarly two candidate RT surfaces for this case, one which remains outside the black hole horizon (analogous to $H_2$ in our discussion) and one which crosses the horizon and ends on the brane (analogous to $\mathcal W$ in our discussion) as pictured in figure \ref{fig:highd}. In the first case the brane is entirely described in the larger boundary region, while in the latter case it is divided into two regions, described in the two parts of the boundary. 

Because the two regions interact, the entanglement entropy is no longer time independent in this case. The discussion in \cite{Cooper:2018cmb} focused on this time dependence; focusing on the entanglement at $t=0$ should be similar to our previous discussion. To clearly identify the portion of the entanglement entropy that is associated with the mixed state on the brane, we would like to work with a solution where the horizon area is large compared to the AdS scale $r_h \gg \ell$, so the entanglement between the brane and the boundary in the semi-classical state is local, as in our discussion. Then to have a large enough region where the surface outside the horizon is minimal, we need the brane to lie far from the horizon, $r_0 \gg r_h$. In \cite{Cooper:2018cmb}, it was found that the brane can't be taken arbitrarily far from the horizon in uncharged black holes, but this is possible if we consider charged black holes close to extremality \cite{Antonini:2019qkt}. 

In this context, of large black holes with the brane far from the horizon, we would expect qualitatively similar results to our analysis: the RT surface which stays outside the horizon will have a portion which lies along the horizon, which we can interpret as giving the entropy $S_{brane}$ of the mixed state on the brane. This will never be more than half the horizon area, and the transition to the RT surface that ends on the brane should occur when $S_{brane}$ is bigger than the area of the surface extending from the horizon to the brane, which should correspond to the boundary term in the island formula from the perspective of the induced gravity theory on the brane.

\section*{Acknowledgements}

SFR thanks Vijay Balasubramanian, Arjun Kar and Tomonori Ugajin for discussions that sparked his interest in this issue. SFR is supported in part by STFC through grants ST/P000371/1 and ST/T000708/1, and SF is supported by an STFC studentship.

\bibliographystyle{utphys}
\bibliography{islands}

\end{document}